\documentclass[%
 preprint,
 amsmath,amssymb,
 aps,
]{revtex4-2}

\usepackage{graphicx}% Include figure files
\usepackage{dcolumn}% Align table columns on decimal point
\usepackage{bm}% bold math
\usepackage{hyperref}% add hypertext capabilities

\begin{document}

\preprint{}

\author{Lan Hai Anh Tran$^*$}
\affiliation{$^*$These authors contributed equally}
\affiliation{%
 School of Chemistry, UNSW Sydney, NSW 2052, Australia}%

\author{Lauren A. Lowe$^*$}
\affiliation{$^*$These authors contributed equally}
\affiliation{%
 School of Chemistry, UNSW Sydney, NSW 2052, Australia}%
\affiliation{%
 Australian Centre for Astrobiology, UNSW Sydney, NSW 2052, Australia}

\author{Yaam Deckel}
\affiliation{%
 School of Chemistry, UNSW Sydney, NSW 2052, Australia}%
\affiliation{%
 Australian Centre for Astrobiology, UNSW Sydney, NSW 2052, Australia}

\author{Matthew Turner}
\affiliation{%
 School of Chemistry, UNSW Sydney, NSW 2052, Australia}%
\affiliation{School of Physics, The University of Sydney, NSW 2006, Australia.}%Lines break automatically or can be forced with \\

\author{James Luong}%
\affiliation{%
 School of Chemistry, UNSW Sydney, NSW 2052, Australia}%
\affiliation{School of Chemistry, The University of Sydney, NSW 2006, Australia.}%Lines break automatically or can be forced with \\

\author{Omar Abdullah A Khamis}
\affiliation{%
 School of Chemistry, UNSW Sydney, NSW 2052, Australia}%

\author{Megan L. Amos}
\affiliation{%
 School of Chemistry, UNSW Sydney, NSW 2052, Australia}%
\affiliation{%
 Australian Centre for Astrobiology, UNSW Sydney, NSW 2052, Australia}

\author{Anna Wang}
\affiliation{%
 School of Chemistry, UNSW Sydney, NSW 2052, Australia}%
\affiliation{%
 Australian Centre for Astrobiology, UNSW Sydney, NSW 2052, Australia}
\affiliation{%
 ARC Centre of Excellence in Synthetic Biology, UNSW Sydney, NSW 2052, Australia}
\affiliation{%
 RNA Institute, UNSW Sydney, NSW 2052, Australia}
\email{anna.wang@unsw.edu.au}

%%%%%%%%%%%%%%%%%%%%%%%%%%%%%%%%%%%%%%%%%%%%%%%%%%%%%%%%%%%%%%%%%%%%%
%% The document title should be given as usual. Some journals require
%% a running title from the author: this should be supplied as an
%% optional argument to \title.
%%%%%%%%%%%%%%%%%%%%%%%%%%%%%%%%%%%%%%%%%%%%%%%%%%%%%%%%%%%%%%%%%%%%%
\title[Holography of vesicles]
  {Measuring vesicle loading with holographic microscopy and bulk light scattering}

% \begin{tocentry}

% Some journals require a graphical entry for the Table of Contents.
% This should be laid out ``print ready'' so that the sizing of the

% Inside the \texttt{tocentry} environment, the font used is Helvetica
% 8\,pt, as required by \emph{Journal of the American Chemical

% The surrounding frame is 9\,cm by 3.5\,cm, which is the maximum
% permitted for  \emph{Journal of the American Chemical Society}
% graphical table of content entries. The box will not resize if the
% content is too big: instead it will overflow the edge of the box.

% This box and the associated title will always be printed on a
% separate page at the end of the document.

% \end{tocentry}

\begin{abstract}
We report efforts to quantify the loading of cell-sized lipid vesicles using in-line digital holographic microscopy. This method does not require fluorescent reporters, fluorescent tracers, or radioactive tracers. A single-color LED light source takes the place of conventional illumination to generate holograms rather than bright field images. By modelling the vesicle's scattering in a microscope with a Lorenz-Mie light scattering model, and comparing the results to data holograms, we are able to measure the vesicle's refractive index and thus loading. Performing the same comparison for bulk light scattering measurements enables retrieval of vesicle loading for nanoscale vesicles.
\end{abstract}

\maketitle

\section{Introduction}
The semi-permeable lipid bilayer membrane is a core feature of all life on Earth~\cite{kindt_bulk_2020}. As a result, entire fields of research are dedicated to lipid bilayer assemblies: they are used as models for plasma membranes~\cite{sarkis_biomimetic_2020, luchini_mimicking_2021, scott_model_2021, simons_model_2004}, bio-mimicking artificial cells~\cite{xu_artificial_2016, siontorou_artificial_2017, rothschild_building_2024, lu_vesicle-based_2022}, and vessels for drug delivery~\cite{liu_review_2022, bozzuto_liposomes_2015, lakshmi_ufasomes_2020}. In order to understand their key function as a biological container, it is critical to have methods of quantifying their loading (i.e. the amount of material they encapsulate), and how that changes as a function of time.

Current techniques such as radiolabelling~\cite{mui_osmotic_1993, anderson_effect_2004, mansy_thermostability_2008, jin_fatty_2018, chen_rna_2005} and fluorescent labelling~\cite{jin_fatty_2018, emami_permeability_2018, chen_rna_2005, mansy_thermostability_2008}, are commonly used for monitoring encapsulated solutes but can be expensive. Moreover, the hydrophobic moieties in fluorescent tags can often interact with the hydrophobic membrane~\cite{hughes_choose_2014} or have undesired interactions with other encapsulated components~\cite{quinn_how_2015}. A label-free technique is thus preferable.

In previous work, we demonstrated that a core-shell light scattering model could be used to measure the thickness of lipid bilayer membranes to within the accuracy of cryo-EM measurements~\cite{wang_core-shell_2019}. This approach required pre-processing the vesicle samples with extrusion through nanometre-sized pores to create a sample of a narrow size distribution and defined (uni)-lamellarity.

\begin{figure}[h]
    \centering
    \includegraphics[width=\textwidth]{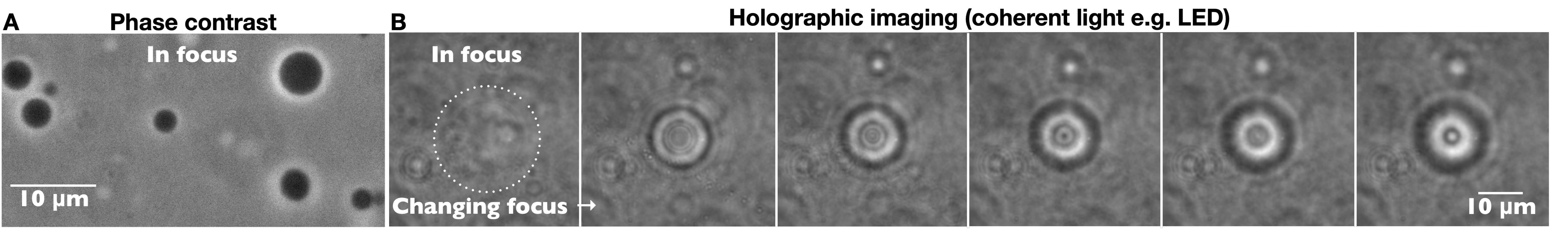}
    \caption{\textbf{A.} Phase contrast and \textbf{B.} holographic images of vesicles encapsulating sucrose (nominal concentration 0.5 M) diluted into an isotonic glucose solution. Under coherent illumination, changing the focal plane of the microscope results in changes in the diffraction pattern of the vesicle. When in focus, the vesicle (inside the white dotted circle) is barely visible in holographic mode.}
    \label{fig:1}
\end{figure}

In this work, we demonstrate a method to determine the loading of single vesicles using light scattering on a minimally-modified microscope. In this technique, known as in-line digital holographic microscopy, a white light source is replaced by a coherent light source, such that the diffraction pattern of the object has more detail in the fringes. Instead of objects becoming blurred when moving out of focus, the hologram arising from interference between undiffracted and diffracted light changes, and provides information about the object's axial position, {which is usually quantified by measuring the object's distance $z$ from the focal plane of the objective} (Fig~\ref{fig:1}). The information contained in the fringes enables objects to be tracked in three dimensions, and for the refractive index and radius to be measured~\cite{lee_characterizing_2007}.

We then fit a generative model for how the objects scatter light to the holograms, using an implementation of Lorenz-Mie scattering within the Python package HoloPy~\cite{barkley_holographic_2020}. Whilst this technique has been used to track biological scatterers such as \textit{E. coli} in 3D~\cite{wang_tracking_2016}, distinguish between populations of scatterers in complex mixtures~\cite{winters_quantitative_2020}, and measure the size and refractive indices of colloidal objects~\cite{martin_-line_2022}, its utility for extracting the refractive index of {vesicles} has not yet been demonstrated.

We find that the solute loading of individual cell-sized vesicles (giant unilamellar vesicles, GUVs) can be quantified from the digital holograms and used to monitor content leakage. From data and modelled holograms, we determine that this technique is optimal for characterising vesicles that have a radius greater than 1 $\mu$m, and a position between 6 $\mu$m and 15 $\mu$m from the focal plane. {This method works well when the solute loading is high enough to achieve a sufficient refractive index contrast with the medium such that the scattered signal is well above the fringe intensities from neighboring vesicles. For weaker scatterers such as vesicles with lower solute loading and much smaller vesicles, we demonstrate that bulk light scattering measurements may be more appropriate.}

\section{Results and Discussions}

\subsection{Modelling}
To extract the refractive index of colloidal objects from holograms, a light scattering model for the assumed geometry of the object/scatterer was used to generate holograms that are iteratively fit to the data hologram. The reasons for using a simple model to fit the hologram are two-fold. First, whilst {a 160 $\times$ 160 pixel hologram of a spherical scatterer takes less than a minute to fit on a typical processor~\cite{dimiduk_random-subset_2014}, the time taken to fit holograms scales with the square of the number of fitting parameters~\cite{fung_imaging_2012}.} {Second, some fitting parameters are strongly correlated and the fitting landscape potentially contains multiple local minima as parameters are adjusted to compensate for each other~\cite{ruffner_lifting_2018, dimiduk_bayesian_2016}. Thus, reducing the number of fitting parameters can aid fit convergence.} We therefore first sought to verify that holograms of vesicles could be fitted effectively with the simplest model -- a homogeneous sphere. 

\begin{figure}[h]
    \centering
    \includegraphics[width=7.5cm]{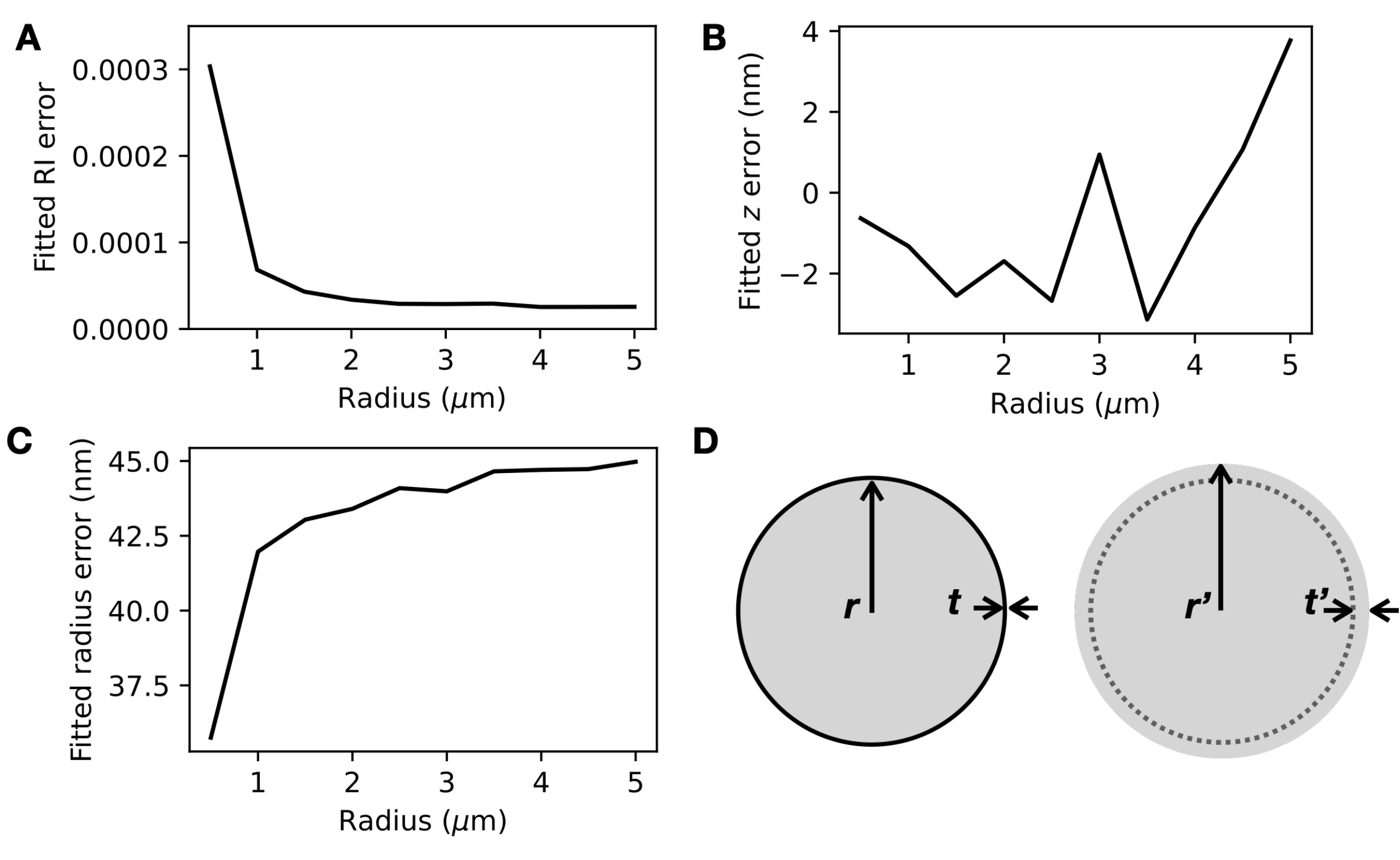}
    \caption{The light scattering model for a homogeneous sphere can be used to extract the refractive index of the contents of a vesicle (a core-shell scatterer with a very thin shell). \textbf{A.} The fitted error in the refractive index $n$ becomes negligible for vesicles larger than 1 $\mu$m in radius. \textbf{B.} The fitted error in $z$ remains below 5 nm and fluctuates with vesicle size. \textbf{C.} The fitted error in $r$ remains approximately 40 nm for all vesicle sizes. \textbf{D.} A schematic showing a core-shell scatterer with inner radius $r$ and shell thickness $t$ (left) and a homogeneous sphere with radius $r'$ = $r$+$t'$ (right).}
    \label{fig:cs}
\end{figure}

Although vesicles are core-shell structures, {with a lipid bilayer corresponding to the shell and the aqueous interior of the vesicle corresponding to the core,} the shell is very thin (approximately 3-5 nm thick~\cite{wang_core-shell_2019}) compared to the typical diameter of vesicles ($\sim \mu$m). Consequently, the shell is expected to contribute far less to the scattering and it may be possible to ignore the presence of the shell in the hologram analysis routine. To test this hypothesis, we used an exact core-shell model for Mie scatterers to model holograms of loaded vesicles (see Experimental) and used a homogeneous sphere model to extract parameters from the holograms. We found that the simple homogeneous sphere model is sufficient for retrieving refractive index information about the vesicle's internal contents: the discrepancy between the fitted refractive index of the vesicle contents and the value used for the core-shell calculation was below 0.0001 refractive index units (RIU) for vesicles larger than 1 $\mu$m in radius (Fig.~\ref{fig:cs}A). The error in the fitted $z$ coordinate less than 5 nm for all vesicle sizes tested (Fig.~\ref{fig:cs}B). {The same trends were seen for vesicles that are bilamellar (Fig.~S1). Whilst the $z$ error appears to increase with vesicle radius, we found that this could be reduced by increasing the analyzed hologram's size (Fig.~S2) to enable more fringes to be analyzed}. 

With the refractive index ($n$) measurement and $z$ localization performing extremely well, the sphere model appeared to compensate for the absence of the shell by fitting to a larger radius $r' = r + t'$, with an error of approximately 42.5 nm (Fig.~\ref{fig:cs}C--D). We suspect that this is because an additional `layer' of vesicle contents could have a similar optical path length to a lipid shell. The optical path length of the additional layer (thickness $t'$ = 42.5 nm, see Fig.~\ref{fig:cs}D) can be calculated by multiplying $t'$ with the layer's refractive index contrast with the medium ($\Delta n \sim$ 0.0077), giving $t' \Delta n \sim$ 0.3279 nm. The optical path length of the lipid shell can be found by multiplying the thickness $t$ = 3 nm with the refractive index contrast with the medium for the lipid, $\Delta n_{lipid} \sim$ 0.1191, which gives $t \Delta n_{lipid} \sim$ 0.3574 nm. The sphere model thus appears to extract $n$ and $z$ information from vesicle holograms well by modelling a slightly larger sphere with a homogeneous refractive index. This is yet another example of the `effective sphere' model working well for inhomogeneous scatterers~\cite{martin_-line_2022, altman_holographic_2021, fung_computational_2019}. 

{One surprising finding was that, given the optimization algorithm used (Levenberg-Marquardt), the homogenous sphere model appeared to be more robust to poor initial guesses than the core-shell models, even when tight constraints were placed on the refractive index $n_{lipid}$ and the thickness of the shell $t$ (Fig.~S3). Another key advantage of using this effective sphere model is that it enables the retrieval of vesicle loading even when the refractive index of the lipid is unknown. Indeed, there are few reports~\cite{khlebtsov_spectroturbidimetry_2001, jones_atmospherically_2015} of lipid refractive index, especially as a function of wavelength.} 

\begin{figure}[h]
    \centering
    \includegraphics[width=7.5cm]{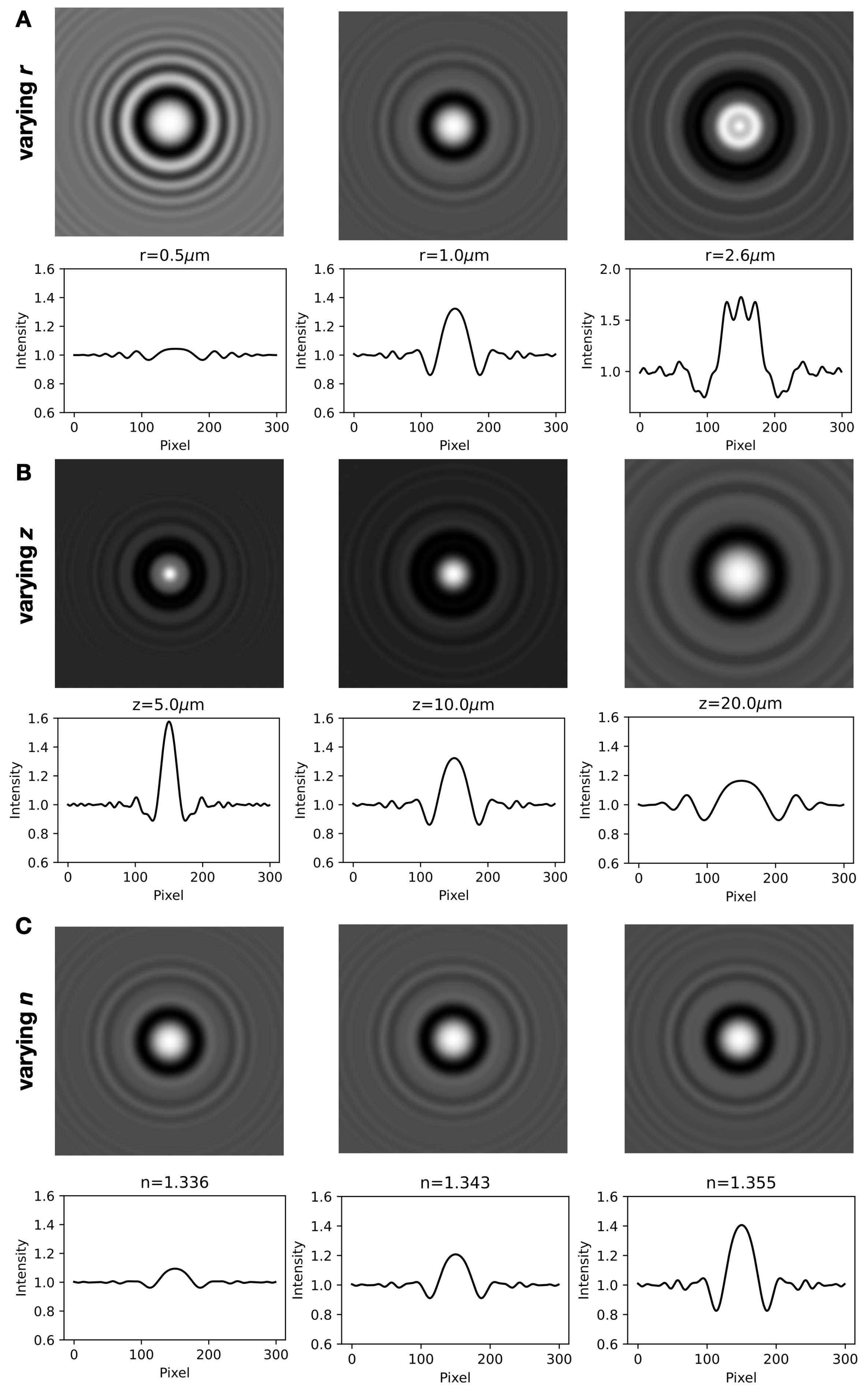}
    \caption{{Holograms and the intensity values across the centre of the hologram were calculated for vesicles with varying \textbf{A.} $r$, \textbf{B.} $z$, and \textbf{C.} refractive index $n$. Varying $r$ and $z$ change the hologram fringe pattern and contrast, whereas varying $n$ only changes the fringe contrast. See also Videos~S1--S3. Parameters used in \textbf{A:} $z$ = 10 $\mu$m, the internal refractive index $n$ = 1.35, the lipid refractive index $n_{lipid}$ = 1.47, and the lipid shell thickness $t$ = 3 nm; \textbf{B:}  $r$ = 1 $\mu$m, $n$ = 1.35, $n_{lipid}$ = 1.47, and $t$ = 3 nm; \textbf{C.} $r$ = 1 $\mu$m, $z$ = 10 $\mu$m, $n_{lipid}$ = 1.47, and $t$ = 3 nm.}}
    \label{fig:nzr}
\end{figure}

We also sought to determine how vesicles of varying sizes, positions, and refractive indices scatter light to gain further insight into the limitations of the technique. As with all Mie scatterers, the fringe pattern and scattering intensity varies non-monotonically with the object's refractive index and size. We found that the holograms contain more features when they are of larger vesicles (Fig.~\ref{fig:nzr}A), and vesicles that are closer to the focal plane of the objective (Fig.~\ref{fig:nzr}B). Two adjustments to image acquisition could therefore improve information retention: a camera with sufficiently small pixel sizes could help to capture the detailed fringe information, and larger image sizes could be used to capture sufficient numbers of fringes. Finally, the refractive index variation is captured in the contrast of the fringes (Fig.~\ref{fig:nzr}C) rather than the fringe pattern or spacing. 

{To analyze the impact of noise on refractive index retrieval, we simulated holograms with different content loadings, with different types of noise (Fig.~S4). Random Gaussian noise did not impact refractive index retrieval by more than 0.0002 RIU (corresponding to $<$5 mM sucrose), even for the weakest-scattering vesicles. Because the information in holograms of spherical objects is radially symmetric, there is a lot of redundant information in holograms and even a small random subset of pixels should contain enough information for retrievals~\cite{dimiduk_random-subset_2014}. Noise taken from experimental holograms, which contain random noise as well as slowly-varying background variations, resulted in errors of no more than 0.0004 RIU (corresponding to $<$10 mM sucrose). The presence of additional fringes from a nearby vesicle impacted the refractive index retrieval more, leading to errors of 0.0008 RIU (corresponding to $<$20 mM sucrose). Very crowded samples thus present the largest challenge for refractive index retrieval, especially for samples that have low solute loading compared to the medium. This is because they have poor hologram fringe contrast and are more easily impacted by the presence of fringes from neighboring vesicles.}

\subsection{Experimental validation}
We opted to use a self-assembly method to encapsulate a model solute, sucrose. This is because methods commonly used to make GUVs that encapsulate a known concentration of solute often require the presence of oil, which can remain as a contaminant in the bilayer~\cite{kirchner_membrane_2012}. While researchers have found that the oil often does not impact the bilayer's mechanical properties, such as rigidity and fluidity, its presence will significantly alter the optical properties~\cite{schaich_characterization_2020}. Furthermore, emulsion-transfer methods can lead to vesicles catastrophically rupturing and thus losing contents. To avoid these complications for this validation study, we followed the protocol for making oleic acid GUVs from micelles as described in detail in Kindt \textit{et al.} and Lowe \textit{et al.}~\cite{kindt_bulk_2020, lowe_methods_2022}, in the presence of 500 mM sucrose. This method has been previously used to encapsulate a range of solutes including small molecule dyes and even colloidal particles~\cite{kindt_bulk_2020, deckel_using_2023}. We then diluted the samples 1 part in 10 into an isotonic solution containing glucose, resulting in vesicles that encapsulated sucrose and maintained a sucrose gradient. 

The vesicles with encapsulated sucrose appear dark under phase contrast imaging as in Figure~\ref{fig:1}A because the glucose solution has a lower refractive index. Holograms of the same vesicle sample are shown in Figure~\ref{fig:1}B. When in focus, the vesicles are almost invisible owing to their low refractive index contrast with the medium. As the focus is shifted, interference fringes appear, revealing information about the contents of the vesicles.

We then fit~\cite{lee_characterizing_2007} a Lorenz-Mie model for how spheres scatter light that takes the objective lens into account~\cite{leahy_large_2020} to the holograms. The input parameters for the model are the vesicle's refractive index $n$, radius $r$, and centroid location {$x$, $y$, $z$}. Examples of best-fit results returned by the Levenberg-Marquardt algorithm are shown in Figure~S5.

\begin{figure}
    \centering
    \includegraphics[width=\textwidth]{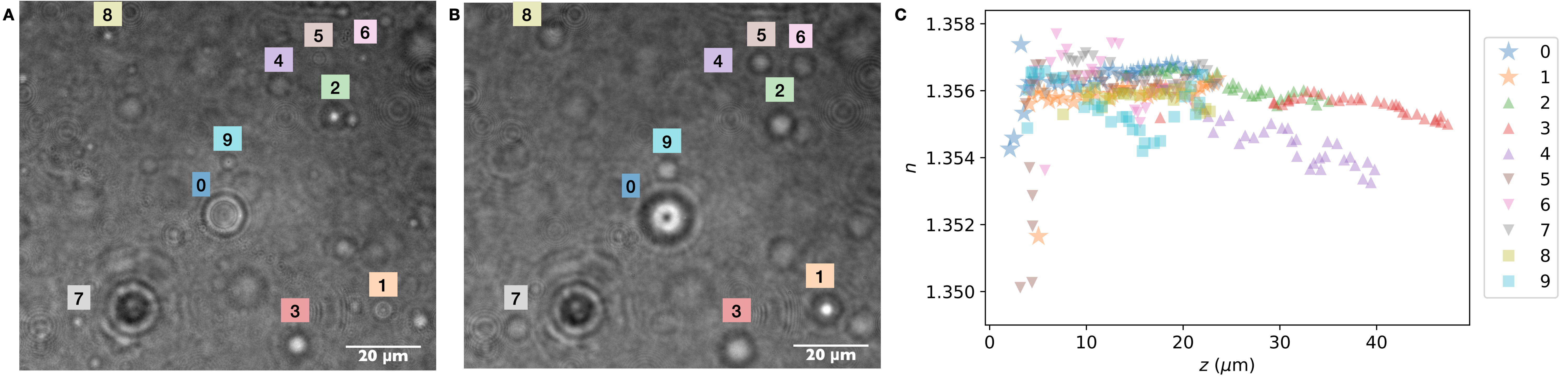}
    \caption{\textbf{A--B.} Holograms of vesicles are captured at several different focal planes. \textbf{C.} The vesicles labelled in \textbf{A--B} were analysed at different $z$ distances to retrieve their refractive index $n$ as a function of $z$. See also Video~S4.}
    \label{fig:scroll}
\end{figure}

Because the detail in the fringes increases with the vesicle's proximity to the focal plane (Fig.~\ref{fig:nzr}B), we sought to determine whether the distance from the focal plane impacted the measured refractive index. Analysing holograms of the same vesicles, but at different focal planes {(whilst allowing $r$ to freely vary during fitting)}, reveals that spherical aberration significantly decreases the measured refractive indices for vesicles within $z < 6~\mu$m of the focal plane, in agreement with the conclusions found by Martin and coworkers~\cite{martin_improving_2021} (Fig.~\ref{fig:scroll}, see also Video~S4). The measured refractive index also decreases with $z > 15~\mu$m, potentially due to poor fringe contrast and interference from nearby objects at these larger distances (see also Fig.~S4). {For vesicles 4 and 9 in Figure~\ref{fig:scroll}, there are fringes from neighboring vesicles visible throughout the hologram series and the degradation in refractive index retrieval is particularly pronounced (see also Fig.~S6).} We therefore recommend taking holograms of vesicles with an axial position $6~\mu$m $< z < 15~\mu$m.

\subsubsection{Measuring vesicle loading}
The most basic use case is to measure the encapsulation of solutes inside vesicles. To correlate a refractive index measurement with vesicle loading, we needed to know the refractive index of sucrose as a function of concentration. We used an Abbe refractometer to measure standard curves for sucrose and glucose solutions in the presence of 100 mM bicine buffer at the sodium line ($\lambda$ = 589 nm; Figure~S7).

\begin{figure}
    \centering
    \includegraphics[width=12cm]{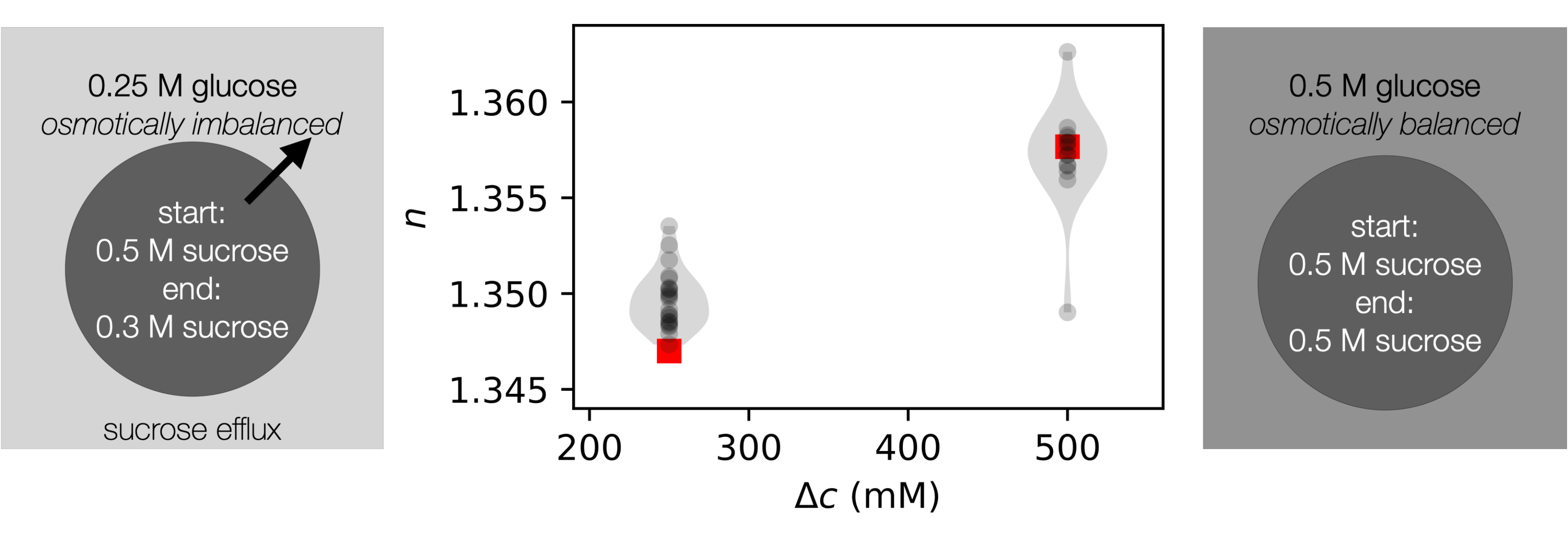}
    \caption{Vesicles encapsulating 500 mM sucrose were diluted into buffers containing either 250 mM glucose (resulting in an osmotic imbalance; {number of vesicles} $N$ = 23) or 500 mM glucose ($N$ = 14). The refractive indices of individual vesicles were measured with holography. For the vesicles exposed to a hypotonic solution, there was content loss of approximately 200 mM sucrose. Refractive indices of 250 mM and 500 mM encapsulated sucrose are shown as red squares. {The vesicles analyzed were between 1--2 $\mu$m in radius.}}
    \label{fig:loading}
\end{figure}

The measured refractive indices reveal that vesicles diluted into an isotonic solution did not exhibit content loss, whereas vesicles diluted into a hypotonic solution did have content loss (Fig.~\ref{fig:loading}). This is in line with expectations of the membrane being semi-permeable; the permeability of water vastly exceeds that of glucose or sucrose, leading to water influx when vesicles are immersed into a hypotonic solution. The strain on the membrane results in rupture, which leads to content loss, membrane resealing, and further cycles of rupture and reseal~\cite{koslov_theory_1984}, until the osmotic stress no longer leads to membrane rupture.

\subsubsection{Measuring vesicle leakage}
Another use case is to measure the leakage of an encapsulated solute over time. We diluted vesicles self-assembled in the presence of sucrose 1 part in 10 into an isotonic solution containing glucose, resulting in vesicles that encapsulate sucrose in the lumen. Because there is both a sucrose and glucose gradient across the membrane, the two sugars are expected to slowly exchange over time, limited by the less permeable solute (sucrose).

\begin{figure}
    \centering
    \includegraphics[width=15cm]{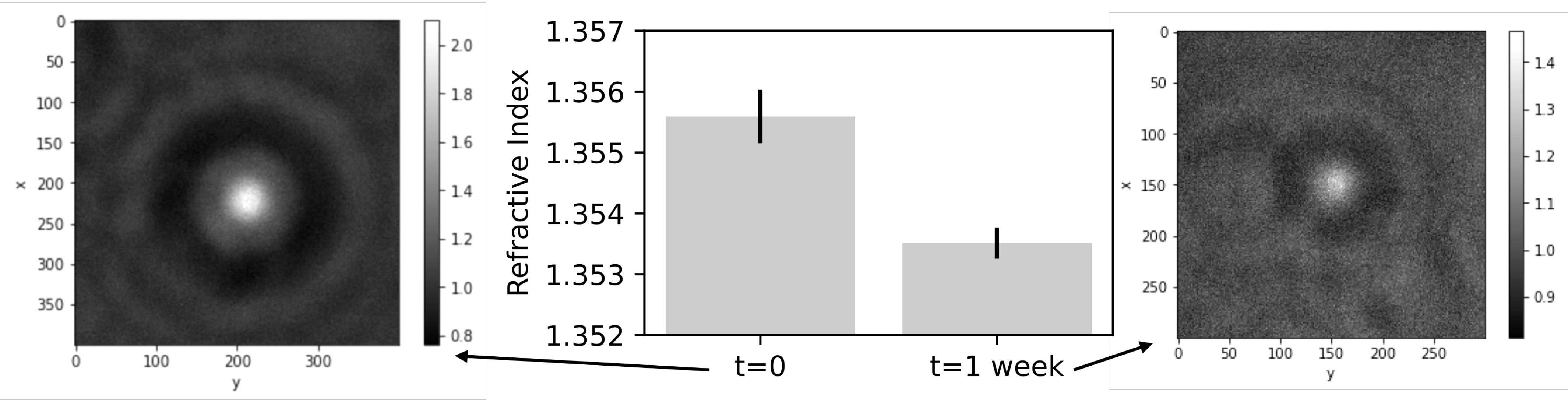}
    \caption{The refractive indices of vesicles encapsulating sucrose in a glucose bath were tracked over one week to determine sucrose/glucose exchange over time ($N$=8). Representative holograms are shown, with dimensions in pixels. Error bars represent the standard deviation from the mean.}
    \label{fig:leakage}
\end{figure}

We found that a sample initially measured $n$ = 1.3556 $\pm$ 0.0004, which corresponds to an encapsulated sucrose concentration of 380 mM (Fig.~\ref{fig:leakage}). Over one week, the same sample had vesicles measuring $n$ = 1.3535 $\pm$ 0.0003, corresponding to the encapsulation of 260 mM sucrose and 120 mM glucose. Given the timescale of a week and the average flux across the membrane, this corresponds to a sucrose permeability of 2 $\times$ 10$^{-11}$ cm/s. This value compares well against the measured permeability of glucose across the same membrane (7 $\times$ 10$^{-11}$ cm/s from Sacerdote and Szostak~\cite{sacerdote_semipermeable_2005}), which is expected to be faster because of its smaller {molecular weight}. 

\subsection{Bulk light scattering measurements}
{The scattering of a single vesicle is challenging to analyse when the refractive index contrast with the medium is sufficiently low, or when the vesicle is sufficiently small. For these cases, we recommend that bulk light scattering measurements (turbidometry) be used to quantify the average vesicle loading in the sample. We previously showed that a core-shell sphere model can be fitted to turbidity measurements of fatty acid and phospholipid vesicle samples to determine vesicle membrane thickness~\cite{wang_core-shell_2019}.}

\begin{figure}[h!]
    \centering
    \includegraphics[width=1\linewidth]{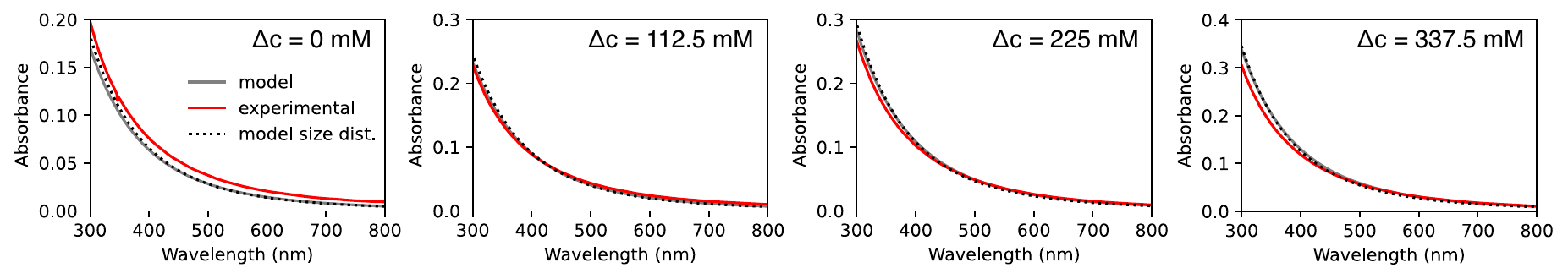}
    \caption{Experimental absorbance spectra (red) of POPC vesicles diluted into isotonic or hypotonic buffers. The modelled absorbance (gray) has no fitting parameters and was determined using the expected sucrose concentration difference ($\Delta{c}$) between the interior and exterior of the vesicles. Taking the vesicle size distribution measured using dynamic light scattering (black, dotted) into account makes little difference to the modelled absorbance.}
    \label{fig:POPC size dist}
\end{figure}

{In Figure~\ref{fig:POPC size dist}, we show the extinction owing to scattering (`Absorbance') of a sample of vesicles encapsulating sucrose, as measured on a UV-visible spectrophotometer. As the concentration difference $\Delta$c between the internal contents and external medium increases, the modelled and experimental absorbance increase.}

{This approach, whilst providing information on smaller vesicles compared to holography, requires that all parameters other than the refractive index of the vesicle's contents $n$ be known and constrained. The most typical method to control for vesicle size involves extruding vesicles through pores, and generates nanoscale vesicles. At these smaller length scales, the exact refractive index and thickness of the membrane all play a large role in the vesicle's scattering, relative to the aqueous core~\cite{wang_core-shell_2019}. However, the refractive index and thickness for most lipid bilayer compositions is unknown. The presence of bilamellar vesicles is also expected to impact the scattering significantly, given the large surface area to volume ratio of these scatterers~\cite{wang_core-shell_2019}. Vesicles prepared via slightly different methods have slightly different distributions in lamellarity, leading to different amounts of sample scattering (Fig.~S8). Turbidometry must therefore be approached with caution, with complementary methods such as cryogenic electron microscopy to constrain the lamellarity, dynamic light scattering to measure the size, and a good estimate for the refractive index and thickness of the lipid, before the turbidity data can be used to extract the refractive index of the vesicles' contents.}

\section{Conclusions}
In summary, we have shown that holographic images of vesicles can be analysed against a Lorenz-Mie light scattering model to quantify the refractive index of the vesicles. The measurement is non-invasive, and requires only microlitres of sample. {The lipid thickness and refractive index does not need to be known if the vesicles are unilamellar. Retrieval of the refractive index $n$ from holograms appears to be robust to within 0.0005 RIU, even in the presence of noise. The main limiting factors for successful $n$ retrieval are the presence of nearby vesicles, and the vesicles being too small. For vesicles smaller than 1 $\mu$m in radius, we demonstrate that bulk light scattering may be more promising under some circumstances}. In future work, this can be expanded to analysing different solutes, and the effects of pores and toxins on the ability of lipid membranes to retain solutes.

\section{Experimental}

\subsection{Chemicals}
Oleic acid ($\geq$99\%), bicine (99\%){, 2-oleoyl-1-palmitoyl-\textit{sn}-glycero-3-phosphocholine (POPC) ($\geq$99.0\%) and chloroform ($\geq$99.8\%) were} purchased from Sigma Aldrich. Sucrose was purchased from Ajax Finechem and D(+)-glucose monohydrate from Calbiochem. 5 M NaOH solution and 10x phosphate-buffered saline (PBS) (1.37 M NaCl, 0.027 M KCl, 0.0147 M {KH$_2$PO$_4$}, 0.081 M {Na$_2$HPO$_4$}) were purchased from Lowy Solutions. All water used was Millipore (18.2 M$\Omega \cdot$ cm). All chemicals were used as received.

\subsection{Vesicle preparation}
Vesicles were prepared by the self-assembly method~\cite{kindt_bulk_2020}. In brief, 5 mM oleic acid vesicles were prepared in a buffer that contained 100 mM Na-bicine (pH 8.3) and up to 500 mM sucrose by adding the appropriate amount of oleate micelles. The microcentrifuge tube was then agitated for 1 week on an orbital shaker at 100 rpm (PSU-10i Grant Bio, UK).

To make 0.1 M oleate micelle stock, 5 M NaOH (30 µL) and oleic acid (31.5 µL) were added to Milli-Q water (970 µL) in a microcentrifuge tube before being placed on an orbital shaker at 100 rpm (PSU-10i Grant Bio, UK) for 1 hour until clear. 1 M bicine stock solution was adjusted to pH 8.3 by the addition of NaOH.

The vesicle suspensions were then diluted ten- to one hundred-fold into a buffer containing 100 mM Na-bicine (pH 8.3) and up to 500 mM glucose. 3 $\mu$L of the diluted vesicle sample was then sealed between a 22 x 22 mm coverslip and a 25 x 75 mm glass slide using silicone vacuum grease (Dow Corning).

\subsection{Imaging}
Vesicles were imaged by phase contrast or holographic modalities using a 1.3 NA 100$\times$ objective (Nikon, Japan) on a TE-2000 inverted microscope (Nikon, Japan). Diascopic illumination was provided by a pT-100 LED (CoolLED, UK). Holographic illumination was provided by a 660 nm mounted LED (Thorlabs, M660 L4, 940 mW, 12 mA, $\lambda$= 660 nm; Thorlabs) following the setup described by Giuliano and coworkers~\cite{giuliano_digital_2014}. Images were captured with a pco.edge 4.2 (PCO Imaging, Germany) using 10 ms exposure time.

\subsection{Bulk light scattering measurements}

{POPC nanoscale vesicles for bulk light scattering measurements were prepared by thin film hydration. 100 $\mu$L of a 100 mM solution of POPC in chloroform was added to a 4 mL glass vial. The sample was heated on a hotplate to remove the solvent and yield a film of POPC. The film was hydrated with 1 mL of 500 mM sucrose in 1x PBS (pH 7.4), and vortexed vigorously for approximately 5 minutes. Samples were sonicated for 1 hour in ice water before being passed 21 times through a polycarbonate filter with pores 100 nm in diameter using a miniextruder (Avanti Polar Lipids). The sample was left to agitate on an orbital shaker (PSU-10i Grant Bio, UK) for at least 1 hour at 100 rpm before being diluted 1 in 10 into a dilution buffer. Dilution buffers were composed of 1x PBS with varying concentrations of sucrose (500 mM, 375 mM, 250 mM or 125 mM) (pH 7.4).}

{The turbidity of extruded vesicle samples was measured using a Jasco V-730 UV-Visible Spectrophotometer and semi-micro UV cuvettes (BRAND), with the dilution buffer used as the blank. Vesicle size was measured with DLS using a Malvern Zetasizer Nano ZS and 12 mm Square Polystyrene Cuvettes (DTS0012) (Malvern Panalytical), with the number averages input into the core-shell sphere model.}

{Bulk scattering calculations were performed using HoloPy as described in Wang \textit{et al.}~\cite{wang_core-shell_2019}, with the inclusion of the concentration of sucrose externally as a known parameter, and the internal concentration of sucrose as a fitting parameter. The additional required parameters were set as follows: radii $r$ as measured using DLS, the lipid refractive index $n_{lipid}$ $\sim$ 1.47~\cite{khlebtsov_studies_2003} with the wavelength dependence as outlined previously~\cite{wang_core-shell_2019}, area per lipid $a$ = 0.627 nm$^2$~\cite{kucerka_fluid_2011}, and lipid shell thickness $t$ = 4.5 nm for POPC.}

\subsection{Hologram calculations and analysis}
Core-shell modelling of vesicle holograms was performed by using the core-shell module in the package HoloPy~\cite{barkley_holographic_2020}. The parameters were set as follows: radii $r$ varying from 0.1 to 5 $\mu$m, $z$ varying from 0 to 20 $\mu$m, and the internal refractive index $n$ varying from 1.3311 to 1.3577. The lipid refractive index was set to $n_{lipid}$ = 1.47, and the lipid shell thickness $t$ = 3 nm.

The holograms were analysed by iterative comparison to a Lorenz-Mie model for scattering from a homogeneous sphere, using HoloPy. As described by Martin and coworkers~\cite{martin_-line_2022}, this procedure can be used to quantify the location, refractive index, and size of the scatterers. In brief, the Lorenz-Mie model is used to calculate the scattered electric field from a vesicle using values for its refractive index, size, and three-dimensional location. This field is then interfered with a plane wave to generate a modelled hologram. The modelled hologram is compared pixel-by-pixel to the data hologram, and the sum of the squared residuals is recorded as the cost function. By continually generating new holograms, a Levenberg-Marquardt algorithm then finds the best-fit values for refractive index, size, and three-dimensional location: the values that minimize the sum of the squared residuals.

The medium index after diluting sucrose-laden vesicles into a glucose medium contained both sucrose and glucose. The refractive index values used for the medium at 589 nm were estimated by linear combinations of the sucrose and glucose values shown in Figure~S7. For hologram fitting, the refractive index of the medium was adjusted to 660 nm by assuming the dispersion of the aqueous medium was dominated by that of water $n_{water}(\lambda)$ = $1.313242+15.7834/\lambda-4382/\lambda^2+1.1455 \times 10^6/\lambda^3$~\cite{quan_empirical_1995, van_engen_dispersion_1998}. The values were then adjusted back to 589 nm for comparison with the measurements from the Abbe refractometer.

For any vesicles with fitted distances closer than $z_{critical}$ = 6 $\mu$m, the spherical aberration was accounted for by adjusting the fitted refractive index values by 0.0009 RIU/$\Delta \mu$m, where $\Delta \mu$m is the difference between the fitted $z$ distance and $z_{critical}$. This slope was determined from fitting the data points with $z <$ 6 $\mu$m in Figure~\ref{fig:scroll}C to a straight line.

\section*{Acknowledgements}
L.A.L. thanks the support of an Australian Government Research Training Program Scholarship. A.W. thanks the support of an Australian Research Council Discovery Early Career Award (DE210100291), Human Frontier Science Program (RGP0029/2020), and support from the Alfred P. Sloan Foundation and Gordon and Betty Moore Foundation.


\bibliography{vesicle-loading}

\end{document}